\documentclass[a4paper,10pt]{article}
\usepackage{graphicx}
\usepackage{subfig}
\usepackage{changepage}
\usepackage{amssymb}
\usepackage{url}
\usepackage{textcomp}
\usepackage{setspace}
\usepackage{sidecap}
\usepackage{scrextend}
\usepackage{xspace}
\usepackage{xcolor,colortbl}
\usepackage{cite}
\usepackage{amsmath}

\usepackage{here,rotating,wrapfig,hyperref,color}

\newcommand\ee{e^+e^-}
\newcommand\aee{A'\to e^+e^-}
\newcommand\xee{X\to e^+e^-}

\usepackage{moreverb} 

\begin{document}

\title{\vspace{5cm}\bf Status and prospects of the NA64 experiment at the CERN SPS \\}

\author{P. Crivelli \\
~\vspace{0.1cm}
ETH Zurich, Institute for Particle Physics and Astrophysics, Auguste-Piccard-Hof 1, 8093 Zurich}
\maketitle

\begin{abstract}
NA64 is a fixed target experiment at the CERN SPS designed as a hermetic general purpose detector to search for Dark Sector physics in missing energy events from electron/positron, hadrons, and  muon scattering off nuclei. In this contribution to the FIPs 2022 workshop, we review the current status and prospects of NA64.
    
\end{abstract}

%

%

\section{The NA64 experiment}

Our proposal (P348) to search for Dark Sectors at the CERN Super Proton Synchrotron (SPS) \cite{Andreas:2013lya} was positively received by the SPS committee (SPSC) in April 2014. We were granted a test beam run in 2015 for a feasibility study, and we were finally approved as the 64th CERN experiment in the North Area (NA64) in March 2016.
NA64 is designed as a hermetic general-purpose detector to search for Dark Sector (DS) physics in missing energy events from electron/positron, hadron, and muon scattering off nuclei. 
The main focus of the NA64 is Light thermal Dark Matter (LDM) interacting with the Standard Model (SM) via vector (or other) portal, such e.g. as dark photons ($A'$) and a variety of New Physics scenarios. 
The experiment, in electron mode (NA64e), employs the optimized  100 GeV electron beam from the H4 beam-line at the North Area. The beam was designed to transport the electrons with the maximal intensity up to a few $\simeq 10^7$ per SPS spill of 4.8 s in the momentum range between 50 and 150 GeV/c.
The hadron contamination in the electron beam was measured to be at a level of $\pi/e^- \lesssim 2\%$ and $K/e^- \lesssim 0.3\%$. \\
The NA64 experiment run from 2016 until 2018, and after the CERN long shutdown (LS2) in 2021, it resumed data taking in a new permanent location at H4 beamline CERN prepared for us. 
Despite the experiment being quite new, very interesting results were rapidly achieved \cite{Banerjee:2016tad,Banerjee:2017hhz,Banerjee:2018vgk}. In this contribution, we review the main results accomplished so far subdividing those into the $A'$ decay modes being explored.

\textbf{Invisible mode:}
NA64 pioneered the active beam dump technique combined with the missing energy measurement to search for invisible decays of massive $A'$, produced in the ECAL target (the electromagnetic (em) calorimeter) by the dark Bremsstrahlung reaction $e^-Z \rightarrow e^-ZA'$, where electrons scatter off a nucleus of charge $Z$.  After its production, the $A'$ would promptly decay into a pair of LDM candidate particles, $A'\rightarrow \chi\chi$, which would escape the setup undetected leaving missing energy as a signature. For this reason, we call these searches {\it invisible}. The parameter space characterized by mixing strengths 10$^{-6} < \epsilon <$ 10$^{-3}$ and masses $m_{A'} $ in the sub-GeV range is the NA64 physics scope: a region where the DM origin can be explained as a thermal freeze-out relic. Missing energy experiments, such as NA64, require precise knowledge of the incoming beam (momentum and particle ID) and an accurate measurement of the deposited energy from the incoming beam's interaction.

\begin{figure}[!htb]
\centering
\includegraphics[width=0.6\columnwidth]{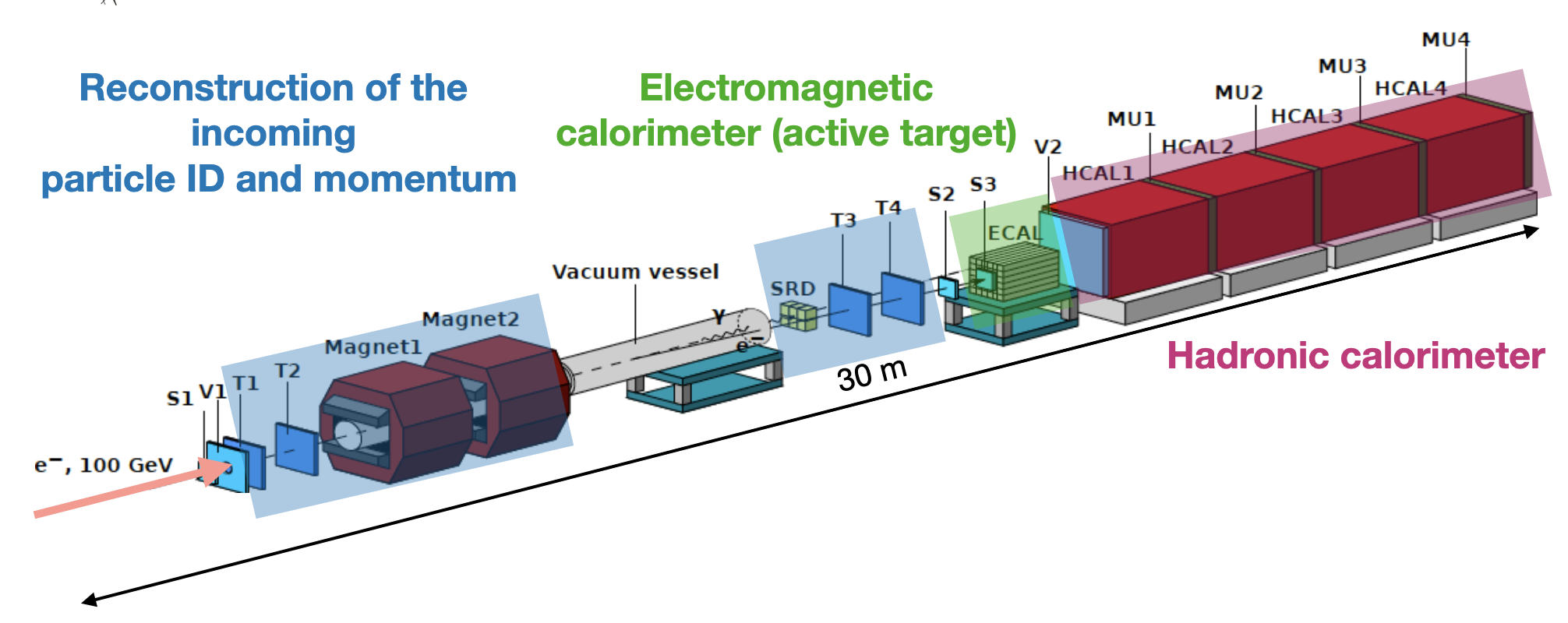}
\caption{NA64 setup and working principle for the search of dark photons through missing energy in the active target (ECAL).}
\label{fig:NA64_sketch}
\end{figure}

A signal event is defined as a single electromagnetic shower in the ECAL with an energy $E_{ECAL}$ below the given threshold\footnote{The value is chosen to maximize the sensitivity of the experiment, in NA64 for 100 GeV incoming beam energy this value is $E_{ECAL}<50$ GeV.} accompanied by a significant missing energy $E_{miss}=E_{A'}=E_{initial}-E_{ECAL}$.  The occurrence of the $A'$ production is inferred in case these events show an excess above those expected from backgrounds. In Fig. \ref{fig:NA64_sketch}, we present a sketch of the setup and a summary of the NA64 working principle.

The signal yield for an active beam approach is proportional to $\epsilon^2$, thus, enhancing the sensitivity for NA64 with respect to the yield  $\propto \alpha_D\epsilon^4$ in traditional beam-dump approach where the $A'$ decay is measured in a detector further away from its production point in the dump.

\begin{figure}[!htb]
\centering
\includegraphics[width=\columnwidth]{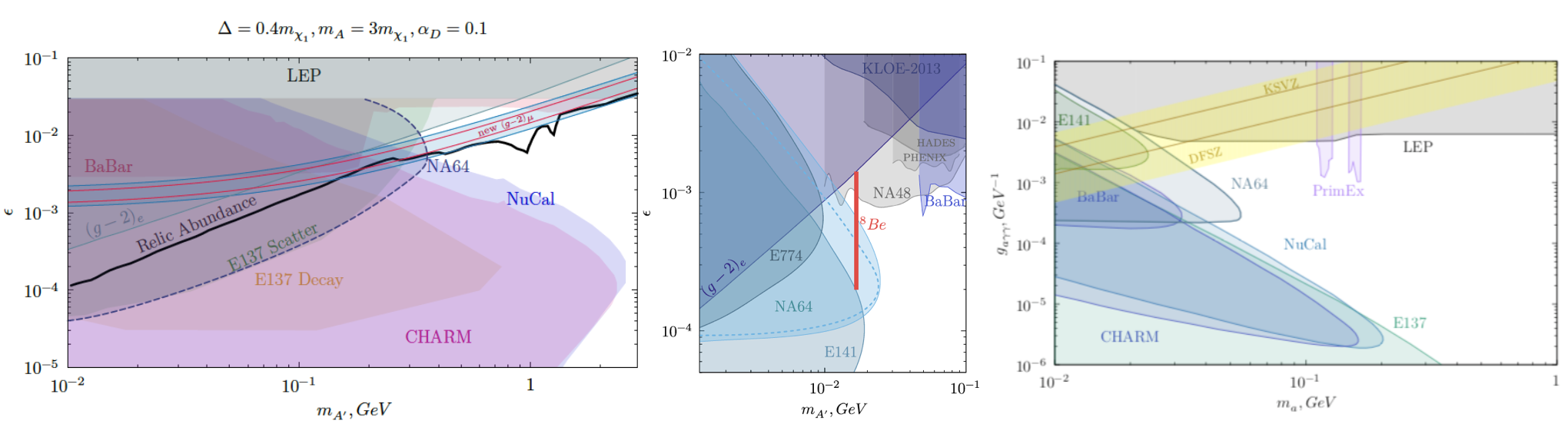}

\caption{Current status of the NA64 experiment 90\% C.L. exclusion limits on semi-visible $A'$ decays  \cite{NA64:2021acr} (Left),  $A'(X)$ visible decays (center) \cite{Banerjee:2019hmi} and NA64 coverage in ALPs searches \cite{Gninenko:2320630} (Right).}
\label{fig:NA64_all}
\end{figure}

The main results achieved so far in the invisible electron mode are:
\begin{enumerate}
\item 2016 July run:  $2.7\times 10^{9}$ electrons on target (EOT) have been collected and the results published in Phys. Rev. Lett. (PRL) \cite{Banerjee:2016tad}. No event compatible with signal was found, thus, excluding most of the favored region of parameter space compatible with the muon $g-2$ anomaly depicted as a band in the left plot of Fig.\ref{fig:NA64_invis}. After our results were published, BABAR completely closed the remaining region of parameter space which could provide an explanation as the Dark photon contributing to the g-2 muon \cite{Lees:2017lec}.
\item 2016 October run: the results corresponding to $4.3\times 10^{10}$ EOT have been published in Phys. Rev. D. \cite{Banerjee:2017hhz}. At this level, NA64 starts to become sensitive to light-dark matter models. No event compatible with a signal was found and so a new exclusion limit could be set.
\item 2017 run: $\simeq 5.5\cdot 10^{10}$ EOT were accumulated and the combined data sample from 2016 and 2017 reaches the milestone of $\simeq 10^{11}$ EOT. 

\item 2018 run: $ 2\cdot 10^{11}$ EOT were collected. No signal-like event was detected. However, the results of the combined analysis from 2016-2018 data, illustrated in Fig. \ref{fig:NA64_invis}, set the most stringent limit for LDM below 0.1 GeV for the canonical benchmark parameters $\alpha_D=0.1$ and $m_{A'}=3m_{\chi}$, thus, NA64 became the leading beam-dump experiment in this region. These results were selected as PRL editor's suggestion \cite{Banerjee:2019pds}.

In addition to the Bremsstrahlung reaction, the resonant $A'$ production channel through the e$^-$ annihilation with the positrons present in the electromagnetic shower has also been considered. The 90$\%$ C.L. exclusion limits from the combined analysis are shown in Fig. \ref{fig:NA64_invis}. The inclusion of the resonant process in the data analysis allows enhancing the NA64 sensitivity for a given dark photon mass resulting in a peak around 200 MeV. The addition of this process improves the NA64 sensitivity in the high mass region, where the Dark photons yield is suppressed due to the $1/m^2_{A'}$ dependency of the Bremsstrahlung cross-section (see \cite{Banerjee:2019pds}). Using positrons as a primary beam instead of electrons would increase by another order of magnitude the sensitivity of NA64 at a given mass depending on the beam energy. By scanning the positron beam energy the mass range probed by this mode can be further expanded. The drawback is that one has to deal with about an order of magnitude more hadron contamination in the beam since the secondary particles are created by the primary 400 GeV SPS protons and thus positively charged hadrons are more abundant than their negative counterpart. To study the impact of the increased hadron contamination and the possible resulting background, a first test beam with 100 GeV positron was taken during the 2022 run (see below).  

Electron/positron beam-dump experiments allow exploring alternative scenarios to the dark photon hypothesis. NA64 has already proven its potential to search for light-scalar and pseudo-scalar axion-like particles (ALPs) produced through the Primakoff reaction \cite{NA64:2020qwq}. The current NA64 coverage in these searches closes part of the gap between beam-dump and LEP bounds and it is shown in the right plot of Fig. \ref{fig:NA64_all}. A search for a generic X-boson coupling to electrons could also be performed. We were positively surprised that the NA64 sensitivity was an order of magnitude more stringent than precision experiments \cite{NA64:2021xzo}. However, one should note that in NA64 we assume the X-boson to decay invisibly while the electron g-2 \cite{Fan:2022eto} and the fine structure measurements \cite{Parker:2018vye,Morel:2020dww} are model-independent. 

\begin{figure}[!htb]
\centering
\includegraphics[width=0.8\columnwidth]{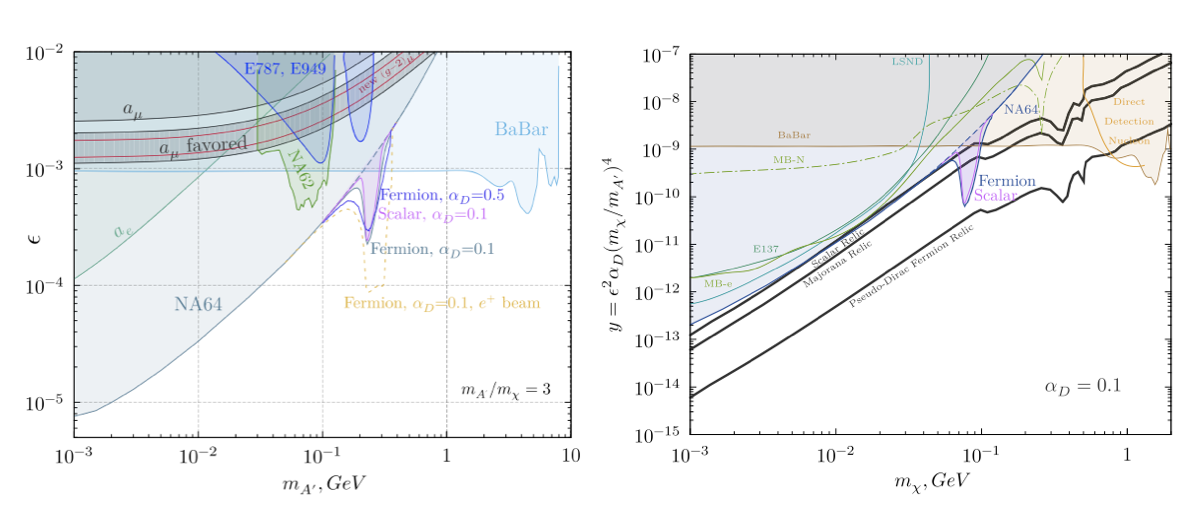}

\caption{Current status of NA64 experiment 90\% C.L. exclusion limits on $A'$ invisible decays including both the Bremsstrahlung and the resonant $A'$ production channels (Left). LDM searches (Right) \cite{Banerjee:2019pds}}.
\label{fig:NA64_invis}
\end{figure}

\end{enumerate}

\textbf{Visible mode:}
 The method for the search of $\aee$ ($\xee$) decays is described in \cite{Andreas:2013lya,Gninenko:2013rka}. 
In this case, the setup is slightly modified to include an additional compact calorimeter upstream with respect to the ECAL. 
If the $A'$ exists, due to the $A'(X) - e^-$ coupling it would occasionally be produced by a shower electron (or positron) in its scattering off a nucleus in the dump: $e^- + Z \to e^- + Z + A'(X)  ;~ A'(X)\to \ee$. Since the $A'$ is penetrating, it would escape the beam dump and
subsequently decay into an $\ee$ pair in a downstream set of detectors. The pair energy would be equal to the energy missing from the target. 
Thus, the signature of the  $A'(X) \to \ee$ decay  is an event with two em-like showers
in the detector: one shower in the dump, and another one in the ECAL, located downstream in this case,  with the sum energy being equal to the beam energy.

\begin{enumerate}
\item 2017 run:  $\simeq 5.5\cdot 10^{10}$ EOT were accumulated. No candidates were found in the signal box. 
The combined 90\% confidence level (C.L.) upper limits for the mixing strength $\epsilon$ were obtained from the corresponding limit for the expected number of signal events. These results set the first limits on the $X-e^-$ coupling in the range $ 1.3\times 10^{-4}\lesssim \epsilon_e \lesssim 4.2\times 10^{-4}$ excluding part of the parameter space suggested by the so called Beryllium anomaly \cite{Krasznahorkay:2015iga} which could be explained by a new X boson with a mass around 17 MeV (named X17)\cite{Feng:2016ysn}. In addition, new bounds are set on the mixing strength of photons with dark photons ($A'$) from non-observation of the decay  $A'\to$ e$^+$e$^-$ of the Bremsstrahlung $A'$ with a mass $\lesssim 23$ MeV. The corresponding paper was highlighted as an editor's suggestion in Phys. Rev. Lett.  5\cite{Banerjee:2018vgk}.

\item 2018 run: about $5\times10^{10}$ EOT were collected at an energy of 150 GeV to boost the putative X bosons outside the calorimeter before it decays in order to improve the sensitivity to higher $X-e^-$ couplings. The results extend the limits to $1.2\times10^{-4}\leq\epsilon\leq6.8\times10^{-4}$ for the vector-like benchmark model and leave only a small region open to fully cover the parameter space compatible with the Beryllium anomaly (see the central panel of Fig. \ref{fig:NA64_all}). The paper was published in Phys. Rev. D Rapid\cite{Banerjee:2019hmi}. Recently, these searches have been extended also to a pseudo-scalar particle decaying visibly into a lepton pair and the result has been published in Phys. Rev. D \cite{NA64:2021aiq}.
\end{enumerate} To completely cover the remaining region of parameter space a new shorter optimized WCAL and a new spectrometer with the possibility to reconstruct the X17 invariant mass should be used as proposed in \cite{NA64:2020xxh}. Everything has been prepared and is ready for installation. However, since it cannot run in parallel with the invisible mode setup we decided to postpone this search. About 30 days of beam-time would be required to solidly probe the remaining X17 parameter space, therefore if the results from PADME currently taking data \cite{Darme:2022zfw}, would confirm this anomaly we would be able to cross-check this in the 2024 run. 

\textbf{Semi-visible mode}: 
Alternative extended scenarios envisioning two DM species split in mass could result in a signature that is a combination of the two signatures described above. A very intriguing feature of this channel is related to the possibility to recover both the DM thermal freeze-out and the $(g-2)_{\mu}$ anomaly explanations, by evading the existing experimental constraints on pure visible and invisible modes \cite{Mohlabeng:2019}. These types of models are known as inelastic DM and we refer to their signatures as a semi-visible channel. An analysis based on a recast of the results from the combined 2016-2018 data \cite{NA64:2021acr} (see the left plot of Fig. \ref{fig:NA64_all}) has already demonstrated the potential of NA64 to study these models. The reach of NA64 to explore in a model-independent way a broad class of parameter space is currently under study (to appear soon on the arXiv).

\section{Current status and prospects of NA64} 
The very dense NA64 program restarted after LS2 in August 2021 with the installation of the setup in the new permanent experimental area in the H4 SPS beam line. The setup was ready for the 6 weeks beam-time in 2021 and took data during 10 weeks in 2022. The goal is to continue our DS exploration until LS3 and collect around $5\cdot10^{12}$ EOT in order to probe the parameter space for light DM models suggested by the observed relic density and other interesting New Physics scenarios. Depending on the results of PADME, in 2024 the upgraded visible setup could be installed to probe the full parameter space of the hypothetical X17 boson.

\begin{figure}[!htb]
\centering
\includegraphics[width=0.3\columnwidth]{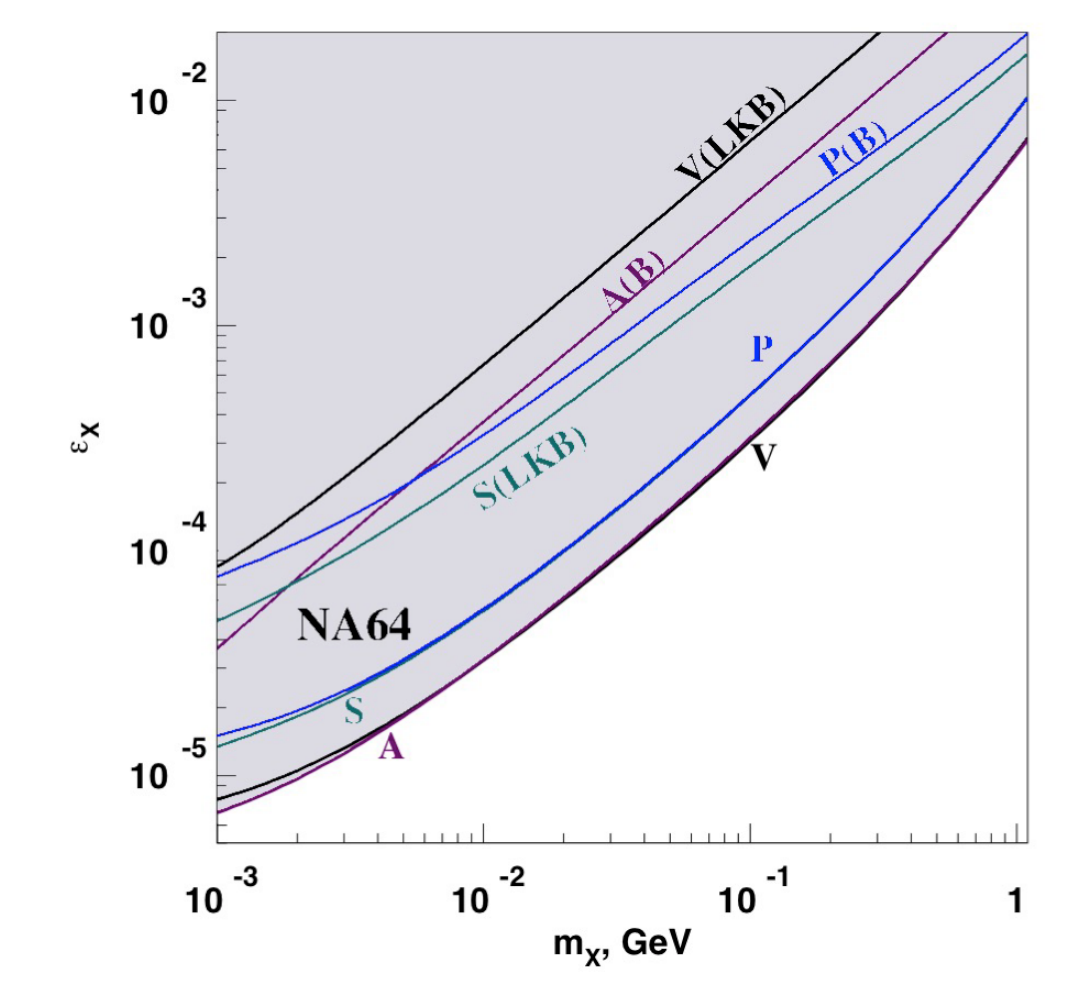}
\includegraphics[width=0.33\columnwidth]{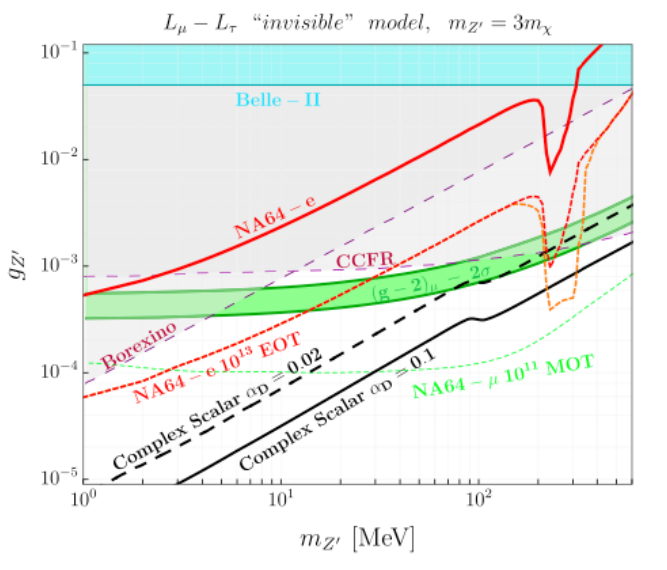}
\includegraphics[width=0.28\columnwidth]{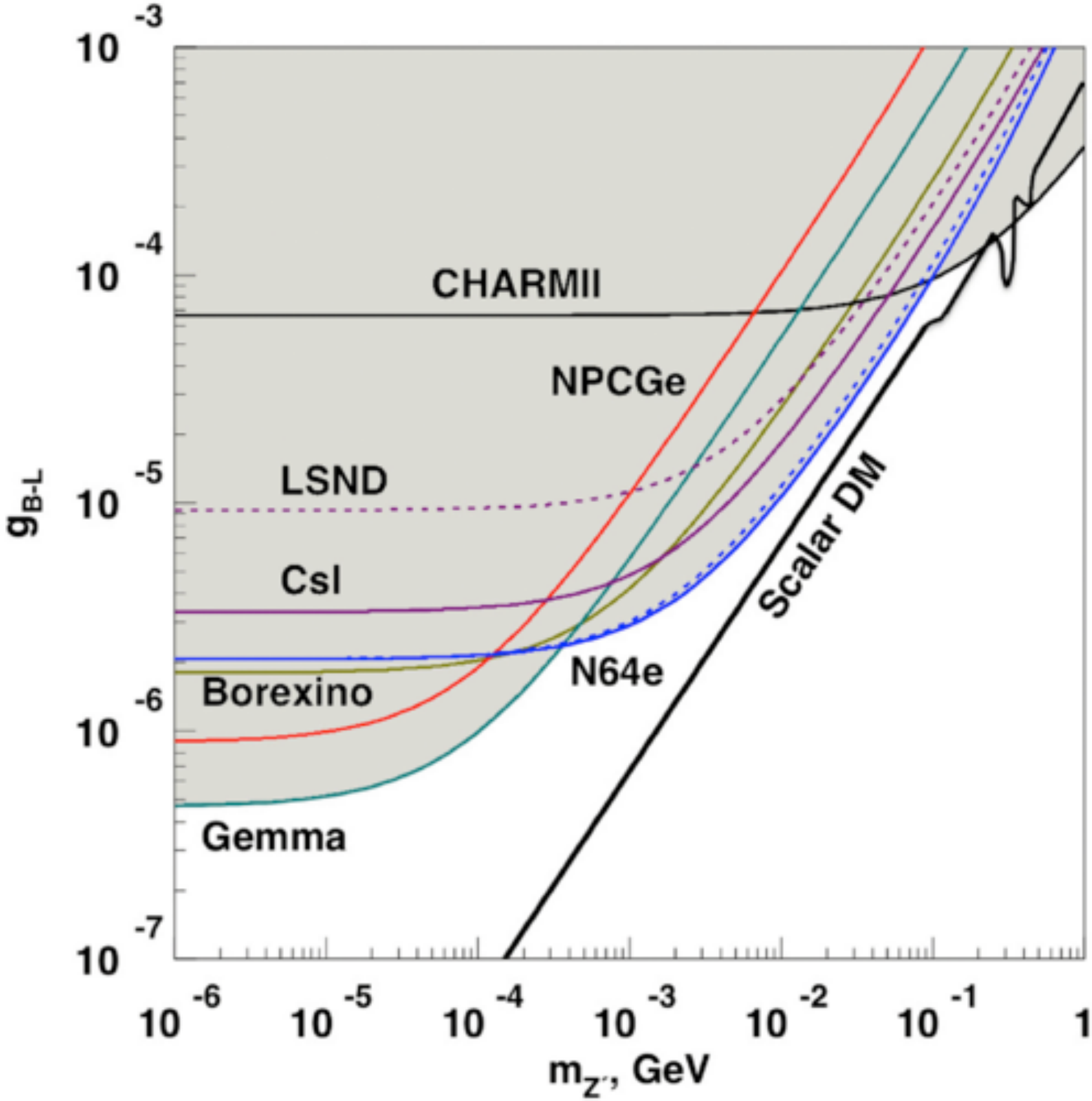}

\caption{Left: NA64 limits for a generic X boson, for the scalar (S), vector (V), pseudo-scalar (P), and axial vector (A) cases \cite{NA64:2021xzo}. Middle: NA64e exclusion limit for the  L$_\mu$-L$_\tau$ (red line) obtained with the 2016-2018 statistics for models where $Z'$ can decay to DM particles \cite{NA64:2022rme}. The red (orange) dashed curves represent the sensitivity projections for a future high-statistics NA64e run with an electron (positron) beam, for a total a of $10^{13}$ EOT, while the green dashed curve is the sensitivity projection of NA64$_\mu$. Right: NA64 exclusion limits for a new B-L Z' boson \cite{NA64:2022yly}.
}
\label{fig:NA64_2022}
\end{figure}

Combining the 2021 data with the one collected before LS2 (total statistic of $3.2\times 10^{11}$ electrons on target), we performed a search for a new $Z'$ gauge boson associated with (un)broken B-L symmetry \cite{NA64:2022yly}. No signal events were found, thus, new constraints on the $Z'$-e coupling strength were set. For the mass range $0.3<m_{Z'}<100$ MeV,  these limits are more stringent than those obtained from the neutrino-electron scattering data (see Fig. \ref{fig:NA64_2022}. The data also indicate that NA64 is background free at a level of $1\times 10^{12}$.
Another possibility that is currently under investigation is based on the existence of a light $Z'$ boson resulting from gauging the difference of the lepton number between the muon and tau flavor. This hypothetical boson can couple via QED vertex corrections to the electron and its existence could explain both the muon g-2 anomaly and the DM relic composition. Moreover, this $Z'$ can be produced again through the dark Bremsstrahlung process, $e^- N \rightarrow e^- N Z'$, but also via the resonant annihilation with secondary positrons from the shower. With the 2016-2018 statistics, NA64 was able to probe in this scenario the region suggested by the $(g-2)_{\mu}$ anomaly up to $m_{Z'}\sim 1$ MeV \cite{NA64:2022rme}.  Such a light $Z'$ can additionally couple directly to muons and its search is therefore also one of the physics goals of NA64$_{\mu}$, the NA64 extension using a high energetic muon beam \cite{Gninenko:2014pea, Sieber:2021fue}.

The NA64$_{\mu}$ program started in 2021 with two pilot runs at the M2 beam-line using the unique CERN 160 GeV/c muon beam.  This will probe DS in a complementary way to the H4 measurements with electrons and will address the g-2 muon anomaly \cite{Sieber:2021fue}.  The main difference between the experimental technique used in NA64$_{\mu}$ compared to NA64e is that in this case one has to rely solely on momentum reconstruction to measure the missing energy carried away from a possible $Z'$ or $A'$ decay. This makes NA64$_{\mu}$ much more challenging than NA64e where one employs calorimeters for this purpose. During the pilot runs in 2021 and 2022, a total of $4\times10^{10}$ MOT were collected. The analysis is still ongoing but the preliminary results already hint to the fact that an additional spectrometer should be added upstream the ECAL since an accurate determination of the incoming momentum is crucial for the experiment. This will be tested in 2023 and the first physics runs are expected for 2024-2025.

It is worth mentioning, that for the new-physics process simulations and for detailed comparison between data and Monte Carlo, a new GEANT4 based package called DMG4 \cite{Bondi:2021nfp} was developed by NA64 members, which has been well accepted by the community, see, e.g \cite{Eichlersmith:2022bit}.

\section{Outlook and conclusions} 
NA64 just reached a major milestone of accumulating $\sim 10^{12}$ EOT which allows one to start probing very interesting LDM benchmark models. The analysis is ongoing and with the increased statistics we expect to improve the sensitivity also of our searches for ALPs, L$_\mu$-L$_\tau$ and B-L $Z'$ bosons, inelastic DM and generic X bosons. The plan until LS3 is to accumulate as many as possible electrons on target and if the background will be under control also use the positron mode to enhance the sensitivity in the higher $A'$ mass region. To study the impact of the larger hadron contamination when running with positrons compared to electrons, in the 2022 run 2 days were used to collect $\sim 10^{10}$ positrons on target (the analysis is ongoing).

NA64 also started its program at the M2 beam-line providing unique high intensity 160 GeV muons to explore dark sectors weakly coupled to muons.  The results of the pilot runs show that with an optimized setup, one could collect  $>10^{11}$ MOT before LS3 in order to check if an L$_\mu$-L$_\tau$ $Z'$ boson is the explanation of the g-2 muon anomaly and complement the searches with electrons (see \cite{Gninenko:2019qiv}). After LS3 the experiment would then continue data taking to accumulate  $\sim10^{13}$ MOT to explore the $A'$ higher mass region and $\mu \to \tau$ and $\mu \to e$ LFV processes \cite{Gninenko:2022ttd}. 

In the 2022 beam-time, we also dedicated 1 day of data taking to accumulate $\sim 2\times 10^{9}$ pions on target in order to understand the potential of NA64 to explore dark sectors coupled predominantly to quarks using the missing energy technique \cite{Gninenko:2014sxa,Gninenko:2015mea}. This will be further investigated and, if the feasibility would be demonstrated, a dedicated search will be performed after LS3. 

To conclude the exploration of the NA64 physics potential has just begun. Our proposed searches with leptonic and hadronic beams provide unique sensitivities highly complementary to similar projects.

\section*{Acknowledgments}
I would like to acknowledge the NA64 collaboration and in particular S. Gninenko, L. Molina-Bueno and V. Poliakov. Special thanks also to the past and current ETHZ members, notably E. Depero, H. Sieber, B. Banto-Oberhauser, M. Mongillo, A. Ponten. A big thank you also to the CERN beam department: D. Banerjee, J. Bernhard, N. Charitonidis,  M. Brugger for their continuous support. My gratitude to the CERN SPSC members and NA64 referees who were fundamental for the success of NA64: particularly C. Vall\'e, G. Lanfranchi, G. Salam, L. Gatignon, M. Wing and G. Schnell.
My work is supported by ETHZ and the Swiss National Science Foundation (SNSF) under the Grants No. 169133 and 186158. 



\bibliographystyle{unsrt}
\bibliography{bibfile}

\end{document}